# Semiconductor to metal transition for Ni doped CdO thin films

Arkaprava Das[a], Ashish Kumar[a], Vijay Soni[a], Soumen Kar[a]

[a] Inter University Accelerator Centre, Aruna Asaf Ali Marg, New Delhi-110067, India.
Email: arkapravadas222@gmail.com

*Abstract*— In the present investigation resistivity Vs temperature measurements have been performed for solgel prepared pure and Ni doped CdO (NDO) thin films. The semiconducting behavior i.e. negative temperature coefficient of resistivity (TCR) of pure CdO thin film through-out the whole temperature range can be attributed mainly due to concentration dependent thermally activation mechanism of the carriers. The semiconductor to metal transition (SMT) for NDO thin films has been elucidated via combined effect of involved grain boundaries, ionic impurities, phonon scattering and carrier activation.

## 1. Introduction

In last few decades carrier transport phenomena in oxide semiconductors has occupied a great attention not only from fundamental aspect but also from optoelectronic device applications. Predominantly the post transition metal oxide like CdO, CdS, CdTe, ZnO, $In_2O_3$ have been thoroughly investigated in last few decades, galvanized by their versatile as transparent conducting oxide (TCOs) in almost the solid state devices and optoelectronics [1]. The essential ingredient for optoelectronic and solid state devices like photovoltaics, energy efficient window, optical wave guide, solar cells, liquid crystal display, gas sensor, electrochromic devices, organic light emitting diodes, ultraviolet semiconductor lasers is TCOs [1]. However there are few challenges or doubts for the selection of the TCOs like chemical stability, cost effectiveness, optical absorption region in the electromagnetic spectrum, conductivity etc. CdO is an n type degenerate TCO semiconductor with a band gap of 2.2 eV having Fermi level and Charge neutrality level above conduction band minima [2]. Considering the aforementioned challenges, CdO overcomes few of them due to its intrinsic properties. It exhibits transparency almost in the whole visible region of the electromagnetic spectrum [3]. It has a high electrical conductivity and high intrinsic mobility without any external doping [2]. Hitherto there is much progress in the research on this conducting oxide but still device parameter for ohmic and Schottky response in CdO/metal heterojunction remains a challenge. Sparse external doping can be used as a tool for creating intentional defects inside the oxide material to tailor its physical properties. Ni has been doped in CdO in the present investigation due to its band gap tunability, low cost, ease of doping and easy availability. Temperature dependent transport properties are very much important to make out the conduction mechanism under divergent condition of doing. At some critical temperature for 1% and 3% Ni doped we observe a semiconductor to metal transition and a change in the temperature coefficient of resistivity which has been explained by the combined effect of involved grain boundaries, ionic impurities, phonon scattering and carrier activation.

## 2. Aim of the study, Experimental method, Results

### 2.1 Aim of the Study

In the present investigation temperature dependent semiconductor to metal transition has been observed within the range of 125 to 140 K for 1% and 3% Ni doped CdO thin films, interestingly which has not been observed for pure CdO thin film. This interesting phenomenon has been explained from the fundamental physics point of view in the present report which might give a new insight into the transport mechanism of carriers for such wide band gap oxide material.

## 2.2 Experimental method

Undoped and doped CdO films were deposited on the corning glass and silicon substrate using sol-gel spin coating technique. Cadmium acetate dihydrate [$Cd(CH_3COO)_2.2H_2O$] and Nickel Acetate tetra-hydrate [$Ni(CH_3COO)_2.4H_2O$] were taken as a source of Cd and Ni in the solution. 2- Methoxyethanol and Di-ethanolamine (DEA) were used as a solvent and sol stabilizer. Solutions were synthesized at room temperature and aged for 2 days. As a precautionary measure the corning glass substrates were rinsed in acetone and deionised water solution in an ultrasonic bath for 10 minutes. The silicon substrates were rinsed in 5% hydrofluoric acid solution (acid + deionised water) in an ultrasonic bath for 10 minutes and subsequently deionised water was flowed to remove the oxidised layer. The films were deposited with the help of a spin coater with a speed of 2800 revolution per minute (rpm) for 30 seconds. The films were dried after each coating using a hot plate at 200 degree centigrade for evaporating organic residuals. This process was repeated for 12 times to get a homogeneous distributed film with desired thickness. The prepared undoped and doped films were annealed at 400 degree centigrade at oxygen environment respectively at a tubular furnace.

The resistivity Vs temperature measurements for both heating and cooling cycle for all the thin films have been performed in the indigenously developed Variable Temperature Insert (VTI) system at the cryogenic department in IUAC New Delhi. The carrier concentration and Hall mobility were measured by Van der Pauw Ecopia HMS-3000 Hall measurement system.

## 2.3 Results

The resistivity Vs temperature measurements for all the thin films have been shown in Fig.1 and values of transport hall measurement parameters have been mentioned in table 1.

The resistivity Vs temperature measurements for all the thin films have been shown in figure 1 and values of transport hall measurement parameters have been mentioned in table 1. The thicknesses of the thin films are 0.2 um determined via Rutherford backscattering. For pure CdO thin film the negative TCR gives the indication of completely semiconducting behaviour through the whole measuring temperature range. However there is little bit hysteresis effect between heating and cooling cycle after 200 k. For pure CdO thin film carrier concentration is very high, therefore increased coulomb screening would reduce the activation energy of native donors [4]. Hence the carrier activation mechanism dominates the semiconducting behaviour. With Ni doping carrier concentration and mobility reduces as conduction band edge shifts towards Fermi stabilization level and formation energy of native donor and acceptor like defects becomes same [5]. With Ni doping grain size reduces as radii of $Ni^{2+}$ and $Cd^{2+}$ are 0.63 Å and 0.97 Å respectively [6]. Usually for n type degenerate semiconductor ionized impurity scattering and grain boundary scattering are expected to be independent of temperature [7]. With reducing grain size the inter-grain scattering is subdued by intra grain scattering mechanism, mainly phonon scattering to influence the transport mobility. For 3% Ni thin film the transition temperature shifts towards higher temperature and overall resistivity increases as the contribution of carrier activation might have increased with increased Ni doping compared to the carrier scattering.

| Sample | Hall mobility ($cm^2/V-s$) | Resistivity (ohm-cm) | Conductivity (Siemens/cm) |
|---|---|---|---|
| p-CdO | 14.13 | 2.78E-03 | 3.59E+02 |
| 1% Ni | 13.82 | 9.30E-03 | 1.07E+02 |
| 3% Ni | 4.57 | 1.59E-02 | 6.28E+01 |

**Table 1.** Room temperature hall measurement data

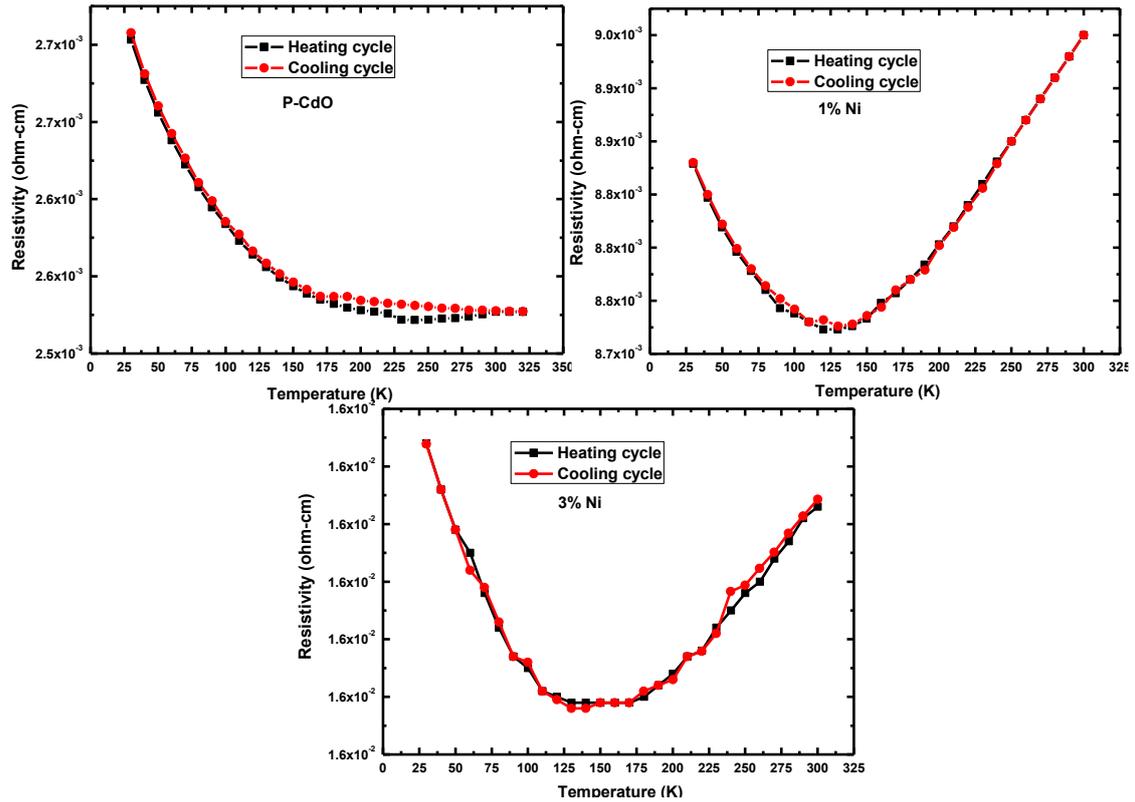

**Fig. 1.** Resistivity Vs Temperature plot for P-CdO, 1% Ni, 3% Ni thin films

**Acknowledgements**

Authors are greatly thankful to the director of IUAC and the cryogenic department of Inter-University Accelerator Centre for the successful transport measurements via indigenously developed Variable Temperature Insert (VTI) system. Authors are also grateful to Sankar Ram Thekkethil and Navneet kumar Suman and Mukesh Kumar for their experimental support. One of the authors (A. Das) acknowledges to Senior Research Fellowship (SRF) Grant Number-F.2-91/1998(SA-1) from University Grant Commission, New Delhi. DST, Govt. Of India is also gratefully acknowledged for granting Science and Engineering Research Board (SERB) project (SB/EMEQ-122/2013).

**References**

[1] C. E. Ekuma, J. Moreno, M. Jarrell, C. E. Ekuma, J. Moreno, and M. Jarrell, "Electronic , transport , optical , and structural properties of rocksalt CdO Electronic , transport , optical , and structural properties of rocksalt CdO," vol. 153705, 2013.

[2] A. Das, S. K. Gautam, D. K. Shukla, and F. Singh, "Correlations of charge neutrality level with electronic structure and pd hybridization," *Sci. Rep.*, vol. 7, 2017.

[3] P. H. Jefferson, S. A. Hatfield, T. D. Veal, P. D. C. King, C. F. McConville, J. Zúñiga–Pérez, and V. Muñoz–Sanjosé, "Bandgap and effective mass of epitaxial cadmium oxide," *Appl. Phys. Lett.*, vol. 92, no. 2, p. 22101, 2008.

[4] A. K. Das, P. Misra, R. S. Ajimsha, A. Bose, S. C. Joshi, D. M. Phase, and L. M. Kukreja, "Studies on temperature dependent semiconductor to metal transitions in ZnO thin films sparsely doped with Al," *J. Appl. Phys.*, vol. 112,

no. 10, p. 103706, 2012.

[5] C. A. Francis, D. M. Detert, G. Chen, O. D. Dubon, K. M. Yu, and W. Walukiewicz, "NixCd1-xO: Semiconducting alloys with extreme type III band offsets," *Appl. Phys. Lett.*, vol. 106, no. 2, pp. 2–6, 2015.

[6] L. Gao, S. Wang, R. Liu, S. Zhai, H. Zhang, J. Wang, and G. Fu, "The effect of Ni doping on the thermoelectric transport properties of CdO ceramics," *J. Alloys Compd.*, vol. 662, pp. 213–219, 2016.

[7] D. H. Zhang and H. L. Ma, "Scattering mechanisms of charge carriers in transparent conducting oxide films," *Appl. Phys. A*, vol. 62, no. 5, pp. 487–492, 1996.